# Reconfigurable Intelligent Surface (RIS) System Level Simulations for Industry Standards

Yifei Yuan, Yuhong Huang, Xin Su, Boyang Duan, Nan Hu, and Marco Di Renzo
China Mobile Research Institute, China
Beijing University of Post-Telecommunication, China
CNRS-CentraleSupélec, Univ. Paris-Saclay (France) and King's College London (UK)

**ABSTRACT**

**Reconfigurable intelligent surface (RIS) is an emerging technology for wireless communications. In this paper, extensive system level simulations are conducted for analyzing the performance of multi-RIS and multi-base-station (BS) scenarios, by considering typical settings for industry standards. Pathloss and large-scale fading are taken into account when modeling the RIS cascaded and direct links. The performance metrics considered are the downlink reference signal received power (RSRP) and the signal to interference noise ratio (SINR). The evaluation methodology is compatible with that utilized for technology studies in industry standards development organizations, by considering the uniqueness of RIS. The simulations are comprehensive, and they take into account different layouts of RIS panels and mobiles in a cell, and different densities and sizes of RIS panels. Several practical aspects are considered, including the interference between RIS panels, the phase quantization of RIS elements, and the failure of RIS elements. The impact of near field effects for the RIS-mobile links is analyzed as well. Simulation results demonstrate the potential of RIS-aided deployments in improving the system capacity and cell coverage in 6G mobile systems.**

## I. INTRODUCTION

Reconfigurable intelligent surface (RIS) is an evolution of multi-input-multi-output (MIMO) antenna technology that benefits heavily from recent breakthrough in the design and manufacturing of metamaterial devices [1]. Via phase, amplitude or polarization tuning of RIS elements, radio environments can be controlled proactively rather than being adapted passively. The favorable conditions created by the deployment of RIS can provide significant performance gains and help to meet the key performance indicators (KPIs) of 6G systems, such as the system spectral efficiency and network energy efficiency. Because of these attractive features, RIS has attracted enormous interests from both academia and industry [2], [3]. Industry groups, such as the IMT-2030 RIS Task Force Group, RIS Tech Alliance (RISTA), and Industry Specification Group (ISG) on RIS within the European Telecommunications Standards Institute (ETSI) have played important roles in promoting RIS standardization and field trials [4]. In 2022, the Third-Generation Partnership Project (3GPP) approved a study item on network-controlled repeaters (NCR) [5] which can be viewed as a prelude for RIS. Apart from wireless communications, RIS can be used in many other scenarios such as sensing, security, energy harvesting, drones, etc. [6], [7]. Due to its large aperture, RIS is expected to benefit from near field propagation as well [8].

Performance analysis of RIS-aided systems can be traced back to early studies on RIS in wireless communications [9], [10]. Since then, many performance evaluation studies have been carried out. For example, a closed-form formula for the weighted sum rate of RIS-aided multi-cell system was derived in [11]. In [12], stochastic geometry analysis was employed to evaluate the data rate of RIS-assisted downlink cellular networks. Via system level simulations, the performance of RIS was assessed in indoor or outdoor environments [13], [14]. As RIS is expected to be a candidate technology for 6G and to be considered by 3GPP in 2025, it is important that the performance of RIS is evaluated by using methodologies and parameters that are compatible to the methods utilized by 3GPP and the International Telecommunication Union (ITU). Yet, such simulations are scanty [15].

In this paper, the same simulation platform as in [15] is utilized, but with significant upgrades. Instead of assuming far field propagation, a near field channel model is considered, which that can model actual RIS-aided environments more precisely, especially the RIS-to-mobile link. The impact of interference modeling on RIS system-level simulations is elaborated. We also study the sensitivity of system-level performance as a function of possible failure of the RIS elements. Overall, the purpose of this paper is to provide a simulation setup for system-level evaluations of RIS-aided communications in multi-cell and multi-RIS mobile networks. Since the performance evaluation is standardization oriented, the simulation methodology and parameters are based on those widely used in 3GPP, with certain modifications that pertain to the deployment of RIS, e.g., channel models for the RIS cascaded link in the near field, and RIS interference modeling. Extensive simulation results to evaluate the downlink reference signal received power (RSRP) and the signal to interference noise ratio (SINR) are provided. The system performance gain provided by the deployment of RIS can serve as a reference for future RIS simulations within 3GPP.

The paper is organized as follows. In Section II, the simulation methodology and system setup are presented, which include the network layout, channel model and RIS model. Simulation results are provided in Section III, where the impact of RIS deployments, near field effects, interference modeling and RIS element failure are elaborated. The paper is concluded in Section IV.

## II. SIMULATION METHODOLOGY AND SYSTEM SETUP

The system-level simulation platform used in this paper is implemented in C++. It is developed based on 5G system simulations that include: 1) network topology; 2) antenna pattern; 3) large/small scale channel model; 4) traffic types and service load models; 5) channel measurement and feedback process; 6) resource scheduling and uplink power control process; 7) SINR statistic, throughput, delay, spectral efficiency and other performance indicators. The architecture is illustrated in Fig. 1 where the legacy parts developed for 5G homogeneous networks are shown in blue background, which include a base station (BS) module, a user equipment (UE, e.g., mobile) module and a link to system level mapping. In each BS or each UE, the functions are divided between the signaling part and traffic part. As for the traffic part, the main processes in terms of system level simulations are concerned with setting the transmitting power, MIMO precoding and SINR calculation. The brown-colored RIS module and cascaded channel module constitute the new components of the system level simulator. In the RIS module, the key processes are beam sweeping and beamforming.

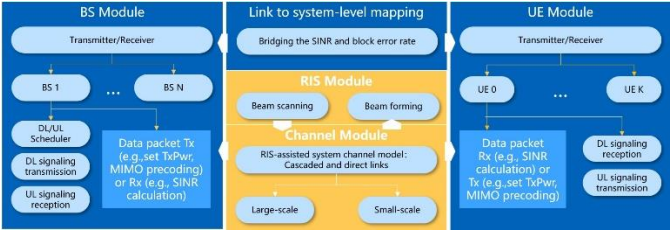

**Figure 1** *Architecture of the RIS system level simulator*

The main simulation parameters are summarized in Table 1 [15].

**Table 1** Simulation set up and parameters.

| Parameter | Value |
|---|---|
| Number of cells | 7 (e.g., 21 sectors), hexagonal macro |
| Operating band | 2.6 GHz |
| Site-to-site dist. | 500 m |
| Number of RIS panels per sector | 4, 8, or 24, min distance of 25 m between RIS panels, uniformly distributed or at cell edges (e.g., 0.9~1.0 cell radius) |
| RIS antenna orientation | Facing towards its serving BS (azimuth) |
| Number of mobiles per sector | 50 (100% outdoor), uniformly distributed or at cell edges (e.g., 0.85~0.9 cell radius) |
| BS antenna height | 25 m |
| BS antenna down-tilt | $0^o$ (mechanical) and $4^o$ (electronic) |
| RIS panel height | 15 m |
| RIS panel down-tilt | $10^o$ (mechanical) |
| Mobile antenna height | 1.5 m |
| BS transmit power | 46 dBm |
| BS antenna gain | 17 dBi for sector beam |
| Polarization | Vertical |
| MS antenna config | $1 \times 2$, with random orientation |
| Pathloss model | ITU-Urban Macro for BS-RIS, RIS-UE and BS-UE links |
| RIS antenna pattern | BS antenna pattern in 3GPP TR 38.901, separately modeled for the cascaded link, total gain of 5 dBi at boresight |
| Number of elements per RIS panel | $16 \times 16$, or $40 \times 40$, with $0.4\lambda$ spacing both vertically and horizontally |
| Number of bits for RIS element phase | 2 bits |
| Combining of RIS cascaded link and direct link | Non-coherent, e.g., gains added up |

### A. Network layout and parameters

The radio access network (RAN) layout comprising of uniform hexagonal cells has been widely used for system level evaluations by standards development organizations (SDOs) such as the 3GPP. In such a layout, each cell, e.g. a BS, has three sectors of equal size but different antenna orientations. The transmit power, antenna gain and antenna height of neighboring macro BSs are usually the same, thus forming a homogeneous network. The wrap-around technique is used to ensure that the cell in the center and the cells in the outer rings see similar interference from neighboring cells. Since the inception of 4G mobile standards, low power nodes (LPN) have been introduced whose transmit power, antenna gains and heights are significantly lower than macro BS, thus forming a heterogeneous network. LPN can serve hot-spots and extend the network coverage. As far the distributed deployment of RIS is concerned, the layout of LPN can be reused.

RIS panels can be deployed at cell edges or uniformly distributed over the network. Similarly, mobiles can be uniformly distributed over the network, or can be located near the cell edges. Similar to system simulations of heterogeneous networks, the number of RIS panels per sector are set to 4, 8 or 16. To get enough statistics in a single drop of UEs, the number of mobiles per sector is set to 50, bigger than the typical value of 20. For simplicity, all users are outdoor. The orientation of RIS panels is important for both the BS-RIS link and RIS-UE link. In the considered system-level simulations, the boresight of each RIS panel points to its serving BS, e.g., the incident azimuth angle of the RIS panel on the BS-RIS link is $0^o$.

In the considered system-level simulations, there can be 19 macro cells or 7 macro cells, corresponding to two rings and one ring of cells. Considering the rather extensive computations needed to model the RIS cascaded link e.g., the large number of RIS elements and near field propagation, the 7 macro cells layout is preferable.

The site-to-site distance between neighboring BSs is set to 500 m which matches the carrier frequency of 2.6 GHz and 46 dBm transmit power in a typical urban macro environment. The antenna height of the BS is 25 m, whereas the heights of the RIS panel and UE are 15 m and 1.5 m, respectively. In light of the relatively large cell size, the mechanical down-tilt and electronic down-tilt of BS antennas are set to $0^o$ and $4^o$, respectively. The mechanical down-tilt of the RIS panel is set to $10^o$. For each sector, there are 4 (horizontal)$\times$8 (vertical) =

32 antenna elements at the BS. The horizontal and vertical spacing between these elements are 0.5λ and 0.8λ, respectively. With this setup, a wide beam can be formed to cover the entire sector with a gain of 17 dBi. For simplicity, only vertical polarized antennas are considered for the BS. Each mobile has two omni-directional antennas, with random orientation.

### B. Channel model

A key difference between the system-level simulations for RIS and LPN is channel modeling. As for LPN with a wired backhaul, only the channel of the access link needs to be modeled. The channel model of the BS-UE link can largely be reused with some adjustments of parameters. As for LPNs with a wireless backhaul, such as relays and repeaters, the backhaul link and the access link can be modeled independently, since the two sides of the antennas are either spatially separated, or operate in time-division multiplexing (TDM) mode. Hence, the channel model for the BS-UE link can mostly be reused with small adjustments. In nearly passive devices like RIS, however, reception and transmission are inherently integrated in a single physical process, e.g., these devices are inherently full-duplex. In other words, an RIS panel is essentially part of the propagation environment and can be considered as an extra boundary condition when a full-wave model is used. Yet, full-wave models seem too complicated for technology studies conducted by SDOs.

The communication model for RIS is somewhat similar to that of an integrated sensing and communication (ISAC) system, in which the target is usually passive. A difference is that an ISAC target may have an irregular shape and a rough surface, while an RIS panel is typically a smooth planar surface. Another difference is that the phase, amplitude or polarization of the RIS elements can be tuned, but those of the sensing targets usually cannot. This means that the channel model of the RIS cascaded link should be partially deterministic so that the model can accurately represent the key mechanism of "controlled anomalous reflection" or, in general, "controlled scattering".

To strike a balance between accuracy and simulation complexity, the channel modeling of the RIS cascaded link can be decomposed into: 1) modeling the propagation channel between the BS and the RIS; 2) modeling the RIS panels; 3) modeling the propagation channel between the RIS and the UE. The propagation channel model can be based on the geometry-based statistical model (GBSM) that is widely used in 3GPP. The modeling of RIS panels involves electromagnetic field analysis.

As for the BS-RIS and RIS-UE links, the GBSM model utilized for the BS-UE in 3GPP TR 38.901 can be used by utilizing appropriate parameters for the RIS, e.g., the height of the RIS panel which is higher than that of mobiles. This results in a higher line-of-sight (LOS) probability and a smaller pathloss of the BS-RIS link, compared to the BS-UE link. This is aligned with the common practice of optimizing the locations of the RIS panels. As for the RIS-UE link, due to the lower height of the RIS panel compared to that of BS antennas, the LOS probability is lower than that of the BS-UE link. However, since the coverage of each RIS panel is relatively small, the probability of having a LOS channel can be quite high for the UE within the RIS coverage region.

To be consistent with the considered site-to-site distance of 500 m, the ITU Urban Macro scenario is assumed for modeling the channel of the BS-RIS link, RIS-UE link and BS-UE link. In each link, the pathloss is generated for each pair of antennas between the transmitter and the receiver if the near field is considered.

### C. RIS model and device parameters

For nearly passive RIS panels, there exist a number of electromagnetic models to characterize the reflection coefficients of the RIS elements in response to external control voltages or currents. Many of these models are quite accurate, however, they are often too complex to be used for system-level simulations by SDOs. By ignoring the detailed physical operation of the RIS, what matters most is the mapping between the tuning signal and the element response. Depending on the type of device, the tuning signal can be the current in a PIN diode, or the voltage in a varactor diode. To reduce the implementation complexity of the control circuits and the cost of the device, the tuning signals have a finite resolution, e.g., they are quantized with a certain number of bits. Apart from factors such as the aperture, material and fine structure of each RIS element, the response is also dependent on the angle of incidence, and additionally on the angle of reflection in some models. In other words, the response of an RIS element is a complex function that can be represented by $f_{id}(\phi_i, \phi_r)$ if two-dimensional beamforming is considered, with $\phi_i$ and $\phi_r$ denoting the angles of incidence and reflection, respectively. The element response depends on the tuning signal which is indexed by "id". In practice, phase tuning is mostly utilized, and this is the case considered in the system-level simulations, reported in the present paper. Hence, an RIS element can be abstracted by an antenna pattern and a tuning signal that controls the phase response, i.e., by a function given by $g_{id}(\phi_i, \phi_r)e^{j\varphi_{id}}$.

Unlike an active element (as usually assumed in 3GPP) whose antenna pattern is only a function of the angle of radiation, the antenna pattern of an RIS element is generally a joint function of the angles of incidence and reflection. This leads to significant changes in current system-level simulations, e.g., the BS-RIS link and RIS-UE link cannot be modeled separately. In the considered system-level simulations, to tackle this issue, the response of an RIS element is simplified by splitting the antenna pattern into the incident component and the reflected component, and the phase tuning is assumed independent of the angles of incidence and reflection, i.e., the response of an RIS element is modeled as $\gamma(\phi_i) \cdot e^{j\varphi_{id}} \cdot \gamma(\phi_r)$. While such simplification may lead to some inaccuracies, its validity was verified in [15]. In the considered system-level simulations, the BS antenna pattern of 3GPP TR 38.901 is used to model $\gamma(\phi)$, but with a total gain slightly lower as compared to the typical value of 8 dBi. Specifically, we consider a value of 5 dBi at boresight.

The number of reconfigurable elements on each RIS panel is assumed to be either $16 \times 16 = 256$ or $40 \times 40 = 1600$. The element spacing is 0.4λ in both the horizontal and vertical directions. Therefore, the size of each RIS panel is about 0.545 m$^2$ or 3.41 m$^2$. The corresponding Rayleigh distances (demarking the far-field region) are 19 m and 118 m, respectively. It is assumed that the phase of the RIS elements is uniformly quantized with 2 bits.

### D. RIS cascaded channel and direct link

Based on the previous section, we can compute: 1) the pathloss of the link between the BS antenna array and each RIS element, 2) the pathloss of the link between each RIS element and the UE antenna, and 3) the amplitude and phase response of the RIS elements. Then, the pathloss of the cascaded link for each RIS element can be calculated by multiplying the obtained three terms and the steering vectors corresponding to the BS-RIS link and RIS-UE link. The pathloss of the cascaded link of the entire RIS panel can be obtained by coherently summing up the contribution of each RIS element. In near-field scenarios, the contributions from different RIS elements are not the same in amplitude, due to the different propagation distances and the different angles of incidence or reflection.

As for the small-scale fading, the RIS cascaded channel can be approximately modeled by considering that the total number of scattering clusters (e.g., paths) is given by the multiplication of the number of paths of the BS-RIS and RIS-UE links. For practical purposes, the total number of considered paths may need to be trimmed to reduce the simulation complexity.

Besides the RIS cascaded link, the direct link (e.g., BS-UE) can be strong, especially in low-frequency bands. Since the RIS is a nearly passive device, the channel coefficients of the cascaded link and direct link should be combined coherently to maximize the received power. In the considered simulation results, however, only large scale fading is modeled and the difference of the propagation delays are not considered either. Hence, the power gains of the RIS cascaded link and BS-UE link are added incoherently in the power domain.

## III. SIMULATION RESULTS

### A. Impact of RIS deployments

Fig. 2a is a heat map showing the geo-distribution of the RSRP gains within the considered 7-cell (21-sector) network. The red dots on the map denote the locations of the RIS panels. The small blue clusters (embedded with sporadic high-power green or yellow spots) represent areas where the deployment of RIS panels results in RSRP gains compared to the baseline system (without RIS), of the order of 5~10 dB.

The cumulative distribution functions (CDFs) of the RSRP are illustrated in Fig. 2b. The UE and RIS panels are distributed near the cell edges. When each RIS panel has $16 \times 16$ elements, the beam-width of the main lobe is about 7.4$^o$ without using special codebooks, which is wide enough to cover the area served by the RIS panel.

Seven CDF curves of the SINR are illustrated in Fig. 2c, as a function of the locations, number and size of the RIS panels. As for the setting with 24 RIS panels each having $16 \times 16$ elements per sector and 4 RIS panels each having $40 \times 40$ elements, the total number of RIS elements per sector is similar, i.e., 6144 vs. 6400. It is observed that, in both cases of uniform distribution of RIS panels and cell-edge deployment of RIS panels, deploying fewer RIS panels of larger size per sector is more preferable than deploying more RIS panels of smaller size per sector. The SINR performance of uniform and cell edge RIS deployments is not very different, except that the SINR is spread more widely when the RIS panels are uniformly distributed. As the number of $40 \times 40$ RIS panels per sector is increased from 4 to 8, the gain in terms of average SINR is increased from 4 dB to 7 dB.

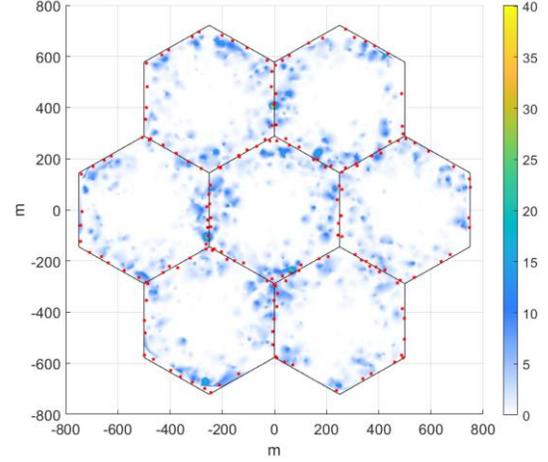

(a) Heat map of RSRP gain

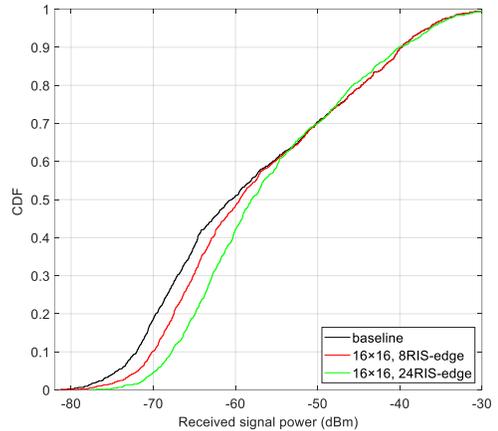

(b) CDFs of the RSRP

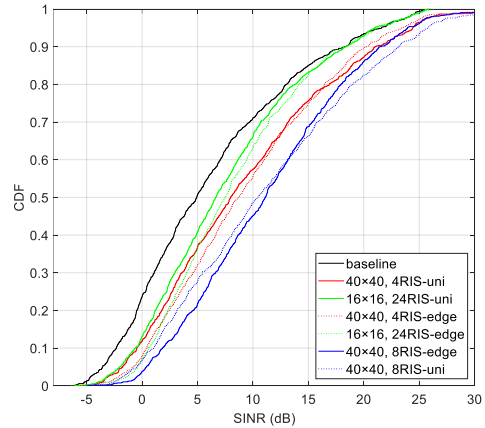

(c) CDFs of the SINR

**Figure 2** *Statistics of downlink RSRP and SINR for different RIS deployments*

## B. Near field analysis

Fig. 3 shows the CDFs of the distance between an RIS panel and the served UEs. The RIS panels are deployed either at the cell edges or are uniformly distributed. The UEs are uniformly distributed. The criterion utilized for the RIS-UE pairing is that the path gain of the BS-RIS-UE cascaded link is no less than 3 dB than that of the BS-UE direct link. When RIS panels with $16 \times 16$ element are deployed, the aperture is 1.0 m. The corresponding Rayleigh distance is roughly 19 m, which is much shorter than the typical distance between a UE and its serving RIS panel as seen in Fig. 3. In other words, the propagation occurs predominantly in the far field. When RIS panels with $40 \times 40$ elements are deployed, however, the aperture is 2.6 m. The corresponding Rayleigh distance is 118 m. According to Fig. 3, there is roughly 50% of probability that a UE would be within 118 m from its serving RIS panel, i.e., the propagation occurs in the near field.

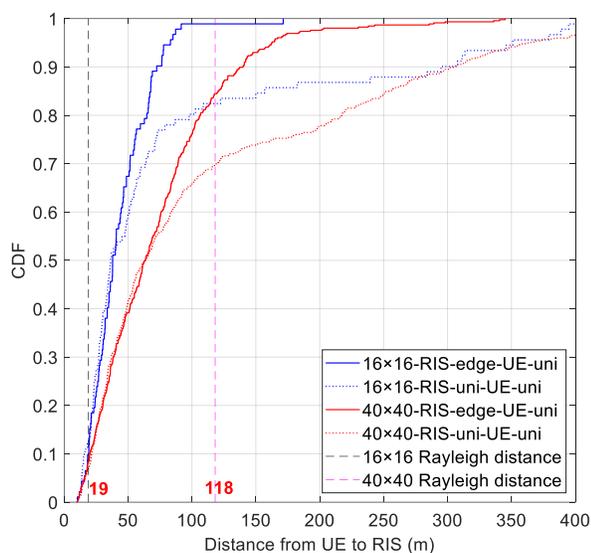

**Figure 3** *CDFs of the distance between an RIS panel and its served UE*

## C. Impact of interference modeling

As a nearly passive device, an RIS cannot distinguish between desired and interfering signals. Hence, an RIS may reflect the interfering signals as the intended signals, which may result in some performance degradation. In the downlink, a UE would see the interference not only from its neighboring BS, but also from the interference that originates from the reflections of RIS panels located in neighboring cells, as well as from the reflections of RIS panels located in its serving sector (for the signals sent by neighboring BS). Fig. 4 compares the SINR distribution by considering the cross-RIS interference and by ignoring it. The RIS panels are located at the cell edges and the UEs are uniformly distributed over the network. It can be seen that the difference is not significant, for the RIS with either $16 \times 16$ elements or $40 \times 40$ elements. Such observation may be explained as follows. When an RIS panel is relatively small, e.g., it has $16 \times 16$ elements, its coverage area is relatively small, e.g., the 50-percentile is 37 m according to Fig. 3. The probability that a UE would be within the coverage area of an RIS panel in neighboring cells or within the coverage area of a non-serving RIS panel in the UE's serving sector is quite small. Hence, these interferences do not seem to be critical in terms of system-level performance. When an RIS panel is relatively large, e.g., it has $40 \times 40$ elements, its coverage area is larger, e.g., the 50-percentile is 59 m based on Fig. 3. Thus, cross-RIS coverage areas may overlap. However, the beamwidth of the RIS panel is rather narrow in this case, e.g., $2.6^\circ$, which implies that the probability of being interfered by a cross RIS panel is still quite small.

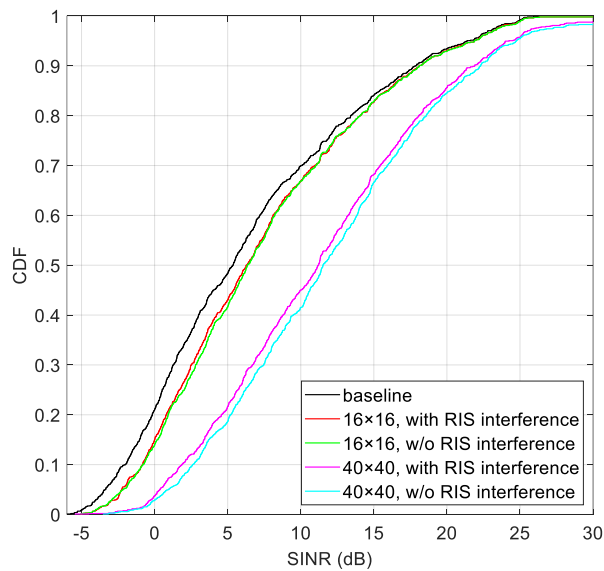

**Figure 4** *CDF of the downlink SINR with and without (w/o) cross RIS interference modeling*

## D. Sensitivity to element failures

Fig. 5 shows the CDF of the downlink SINR when the percentage of failure of the RIS elements is 0%, 5% and 10% for RIS panels with $16 \times 16$ and $40 \times 40$ elements. In the considered study, the RIS element failure is modeled as a random phase of 2 bits, meaning that the phase shift applied by the RIS element is out of control. The location of the malfunctioning RIS elements is also random, e.g., it is uniformly distributed over the whole RIS panel. It can be observed that the impact of RIS elements failure is rather small, as long as the failure rate is below 10%. Such finding applies to both relatively small RIS panels, e.g., $16 \times 16$ elements, and relatively large RIS panels, e.g., $40 \times 40$ elements. The insensitivity to failures of the RIS elements may be explained by the embedded plot in Fig. 5, where the beam patterns corresponding to the 0%, 5% and 10% failure rates are compared. It can be seen that although the presence of malfunctioning RIS elements increases the sidelobes, the main lobe is still 7 dB stronger than the highest sidelobe. Also, there is no shifting in the beam directions compared to the ideal case of absence of malfunctioning RIS elements. Considering that the use of a finite phase tuning, e.g., 2 bits, increases the sidelobes in any case, the presence of some additional small sidelobes does not significantly degrade the system-level performance.

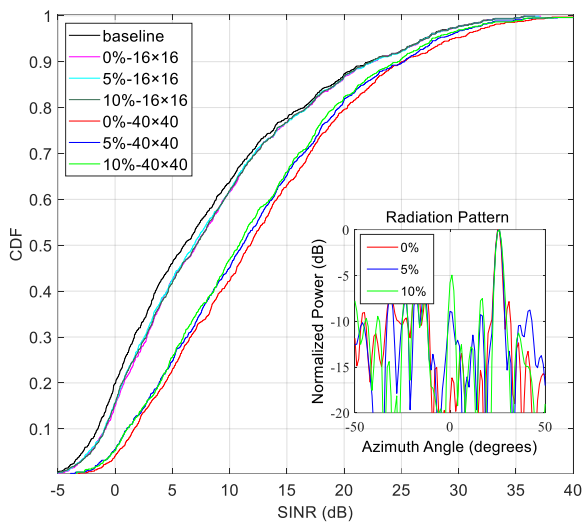

**Figure 5** *Sensitivity of system-level performance to RIS element failures*

## IV. SOME INSIGHTS

While the reported system-level simulations are rather preliminary, some important insights can be drawn from the obtained results which may be useful for future studies and deployments of RIS.

As a nearly passive device without power amplifiers, the effectiveness of an RIS panel is fundamentally determined by the total aperture of the panel. Whether it is more preferrable to deploy more panels of smaller sizes or fewer panels of larger sizes depends on the network layout, channel model, etc. In the considered typical 3GPP setting operating in the 2.6 GHz band, deploying fewer RIS panels of larger size seems to be a better choice. This means that more efforts should be invested to design larger RIS panels, rather than searching for suitable sites for deploying such panels. In the mid-band utilized by future 6G mobile systems, e.g., 6.7 GHz, the size of RIS panels with $40 \times 40$ elements spaced half wavelength apart is roughly $0.9 \times 0.9$ m$^2$, which may not be too large for realistic deployments. When the RIS element spacing is fixed, more RIS elements are needed to build larger panels, which would increase the hardware cost, e.g., more diodes and more complicated tuning circuits. The use of a wider RIS element spacing, e.g., larger than half a wavelength, may be considered to alleviate the implementation burden, while still benefiting from the aperture gain. Implementations based on metagrating designs may also be considered, for ensuring larger spacings among the RIS elements and hence easing the implementation of control circuits on the RIS panels.

One challenge when manufacturing large RIS panels is to ensure that most RIS elements work properly. Multiple iterations are be required for quality control, which increases the device cost. From the obtained preliminary system-level simulation results, it seems that a failure rate of about 10% can be tolerated, which implies that the quality control for RIS manufacturing may be relaxed. This means that RIS panels may not be very sensitive to the aging of the RIS elements, so that an RIS can operate in harsh propagation environments for a relatively long time.

As the aperture of the RIS panel increases, near field propagation becomes prevailing for the RIS-UE link. The codebooks utilized for far field propagation would be too sub-optimal in this setting, and new codebook designs for the phase shifts of the RIS elements, channel state information (CSI) estimation and feedback, etc. are necessary.

The interference generated by RIS panels may not be as significant as it may appear at the first sight. To compensate for the multiplicative pathloss on the cascaded RIS-aided links, an RIS panel needs to form narrow beams towards desired directions of reflection. For larger RIS panels, narrow beams will be widely used. Hence, the probability of interfering other network nodes would be relatively low. These considerations apply to the analysis of coexistence scenarios where the operating bands of different operators are adjacent to one another (inter-operator interference). Since an RIS usually does not have filters to limit the radiated signals within the operating band, the deployment of an RIS may potentially cause involuntary interference to adjacent bands. However, because the beams are rather narrow, the probability of being "hit" by an interfering beam seems low.

## V. Conclusions

In this paper, system-level simulation results for RIS-aided cellular networks were presented, by considering an evaluation methodology and parameters that are suitable for technology studies utilized by SDOs such as the 3GPP. System level simulations were carried out considering various RIS layouts, sizes and numbers of RIS panels. Significant gains in terms of downlink RSRP and SINR are observed as the size of the RIS panels increases. Simulation results revealed the existence of near field propagation for the RIS-to-UE links. The obtained simulation results also reveal that the impact of interference and RIS element failure is not very significant, which is a promising outcome toward the wide deployment of RIS panels in realistic 6G cellular networks.

## References

[1] M. Di Renzo, A. Zappone, M. Debbah, et. al., "Smart radio environments empowered by reconfigurable intelligent surfaces: how it works, state of research and the road ahead," *IEEE Journal of Sel. Areas in Comm.*, vol. 38, no. 11, Nov. 2020, pp. 2450-2525.

[2] Y. Yuan, Q. Gu, A. Wang, et. al., "Recent progress on research and development of reconfigurable intelligent surface," *ZTE Communications*, vol. 20, no. 1, Mar. 2022, pp. 3-13.

[3] R. Liu, Q. Wu, M. Di Renzo, et. al, "A path to smart radio environments: an industrial viewpoint on reconfigurable intelligent surfaces," *IEEE Wireless Comms*, vol. 29, no. 1, Feb. 2022, pp. 202-208.

[4] J. Sang, Y. Yuan, W. Tang, et. al., "Coverage enhancement by deploying RIS in 5G commercial mobile networks: field

trials," *IEEE Wireless Comm.*, vol. 31, no. 1, Feb 2024, pp. 172-180.
[5] 3GPP, RP-222673, New WID on NR network-controlled repeaters, ZTE, RAN#97-e, Sept., 2022.
[6] W. Shi, W. Xu, X. You, et. al., "Intelligent reflection enabling technologies for integrated and green Internet-of-Everything beyond 5G: Communication, sensing, and security," *IEEE Wireless Communications,* vol. 30, no. 2, Apr. 2023, pp. 147-154.
[7] E. Basar, G. C. Alexandropoulos, Y. Liu, et. al., "Reconfigurable intelligent surfaces for 6G: emerging hardware architectures, applications, and open challenges," *IEEE Veh. Tech. Magazine,* vol. 19, no. 3, Sept 2024, pp. 27-47.
[8] J. An, C. Yuen, L. Dai, et. al, "Near-field communications: research advances, potential, and challenges," *IEEE Wireless Comm.*, vol. 31, no. 3, June 2024, pp. 100-107.
[9] Q. Wu and R. Zhang, "Beamforming optimization for wireless network aided by intelligent reflecting surface with discrete phase shifts," *IEEE Trans. on Comm*, vol. 68, no. 3, March 2019, pp. 1838–1851
[10] M. Di Renzo, K. Ntontin, J. Song, et. al., "Reconfigurable intelligent surfaces vs. relaying: differences, similarities, and performance comparison," *IEEE Open J. Commun. Soc.*, vol. 1, 2020, pp. 798-807.
[11] C. Pan, H. Ren, K. Wang, et. al., "Multi-cell MIMO communications relying on intelligent reflecting surfaces," *IEEE Trans. on Wireless Comm.*, vol. 19, no. 8, Aug. 2020, pp. 5218-5233.
[12] T. Shafique, H. Tabassum, and E. Hossain, "Stochastic geometry analysis of IRS-assisted downlink cellular networks," *IEEE Trans. on Comm.*, vol. 70, no. 2, Feb. 2022, pp. 1442-1456.
[13] I. Yildirim. A. Uyrus, and E. Basar, "Modeling and analysis of reconfigurable intelligent surfaces for indoor and outdoor applications in future wireless networks," *IEEE Trans. on Comm.*, vol. 69, no. 2, Feb. 2021, pp.1290-1301.
[14] B. Sihlbom, M. I, Poulakis, M. Di. Renzo "Reconfigurable intelligent surfaces: performance assessment through a system-level simulator," *IEEE Wireless Comm.*, vol. 30, no. 4, Aug. 2023, pp. 98-106.
[15] Q. Gu, D. Wu, X. Su, et. al., "System-level simulation of RIS assisted wireless communications system," *IEEE Proc. of Globecom* 2022, pp. 1540-1545.

## Biographies

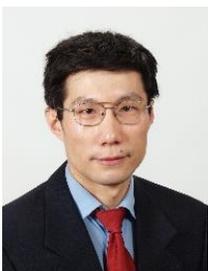

Yifei Yuan (Fellow, IEEE) was with Alcatel-Lucent and ZTE Corporation He joined China Mobile Research Institute in 2020 as a Chief Expert, responsible for 6G air interface study. He has extensive publications, including 13 books on LTE-Advanced, 5G and 6G. He has over 60 granted US patents.

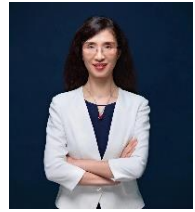

Yuhong Huang is the General Director of China Mobile Research Institute. Over past 30 years, she led many key projects of 2G till 6G.

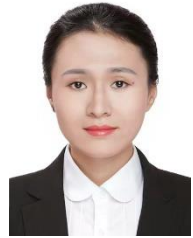

Xin Su is a Senior Engineer at China Mobile Research Institute.

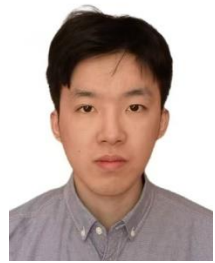

Boyang Duan is a master student at Beijing University of Post-Telecommunication.

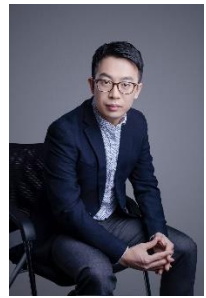

Nan Hu is a vice department head at China Mobile Research Institute and the vice chair of 3GPP RAN Technical Specification Group.

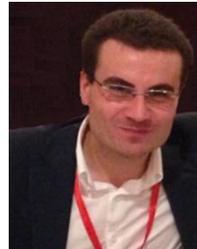

Marco Di Renzo (Fellow, IEEE) is a CNRS Research Director (Professor) and the Head of the iPhyCom group, Laboratory of Signals and Systems, Paris-Saclay University, as well as a Chair Professor with the Centre for Telecommunications Research - King's College London, London. He is a Highly Cited Researcher and is currently serving at the Director of Journals of the IEEE Communications Society. He is a co-founder, vice-chair and rapporteur of ISG-RIS within ETSI.